\documentclass{article}
 \usepackage{amssymb}
 \usepackage{amsmath}
 \begin{document}
 \def\Ref#1{(\ref{#1})}
 \def\p{\partial}
 \def\be{\begin{equation}}
 \def\bea{\begin{eqnarray}}
 \def\ee{\end{equation}}
 \def\eea{\end{eqnarray}}
 
 \noindent{\large\bf Discrete scale invariance, and its
 logarithmic extension}
 
 \vskip 2\baselineskip\noindent
 N. Abed-Pour$^{a,1}$, A. Aghamohammadi$^{b,d,2}$, M.
 Khorrami$^{c,d,3}$,\\
 \& M. Reza Rahimi Tabar$^{a,d,e,4}$\\
 \\
 \begin{itemize}
 \item[$^a$]{\it Dept. of Physics, IUST, P.O.Box
 16844, Tehran, Iran}
 \item[$^b$]{\it Department of Physics, Alzahra
 University,
           Tehran 19834, Iran}
 \item[$^c$]{\it Institute for Advanced Studies in
 Basic Sciences,
           P. O. Box 159, Zanjan 45195, Iran}
 \item[$^d$]{\it Institute for Applied Physics, P. O.
 Box 5878,
           Tehran 15875, Iran}
 \item[$^e$]{\it CNRS UMR 6529, Observatoire de la
 C\^{o}te d'Azur,
           BP 4229, 06304 Nice Cedex 4, France}
 \end{itemize}
      $^1$ abed-pour@iust.ac.ir\\
      $^2$ mohamadi@azzahra.ac.ir\\
      $^3$ mamwad@iasbs.ac.ir\\
      $^4$ rahimitabar@iust.ac.ir
 
 \begin{abstract}\noindent
 It is known that discrete scale invariance leads to
 log-periodic
 corrections to scaling. We investigate the
 correlations of a
 system with discrete scale symmetry, discuss in
 detail possible
 extension of this symmetry such as translation and
 inversion, and
 find general forms for correlation functions.\\
 \\
 PACS: 11.10.-z, 11.25.Hf
 \end{abstract}
 \newpage
 \section{Introduction}

 Log-periodicity is a signature of {\em discrete}
 scale invariance
 (DSI) [1]. DSI is a symmetry weaker than
 (continuous) scale
 invariance. This latter symmetry manifests itself as
 the
 invariance of a correlator ${\cal O}(x)$ as a
 function of the {\em
 control} parameter $x$, under the scaling $x\to
 e^\mu x$ for
 arbitrary $\mu$. This means, there exists a number
 $f (\mu)$ such
 that
 \begin{equation}\label{i1}
 {\cal O}(x) = f(\mu){\cal O}(e^\mu x).
 \end{equation}
 The solution to \Ref{i1} is simply a power law
 ${\cal O}(x)=
 x^{\alpha}$, with $\alpha =-(\log f)/\mu$, which can
 be directly
 verified. In a system having DSI, correlators obey
 scale
 invariance, only for {\em discrete} values of the
 magnification
 factor $\mu=n\mu_1$, where $n$ is integer and
 $\mu_1$ is fixed.
 $\mu_1$ is the period of the resulting
 log-periodicity \cite{1}.
 This property can be qualitatively seen to encode a
 {\em
 lacunarity} of the fractal structure. The most
 general solution to
 \Ref{i1} with $\mu=n\mu_1$ is
 \begin{equation} \label{i2}
 {\cal O}(x) = x^{\alpha}~P\left({{\ln
 x}\over{\mu_1}}\right),
 \end{equation}
 where $P(y)$ is an arbitrary periodic function of
 period $1$,
 hence the name log-periodicity. Expanding it in
 Fourier series
 $\sum_{n=-\infty}^\infty  c_n \exp[(2n\pi i(\ln
 x)/\mu_1])$, we
 see that ${\cal O}(x)$ becomes a sum of power laws
 with the
 exponents $\alpha_n=\alpha+(2n\pi i/\mu_1)$, where
 $n$ is an
 arbitrary integer. So there is an infinite set of
 discrete complex
 exponents. Specifically, it has been established
 that for
 financial bubbles prior to large crashes, a first
 order
 representation of \Ref{i2},
 \begin{equation}
 I(t)=A+B(t_c-t)^\beta+C(t_c-t)^\beta\cos(\omega\ln(t_c-t)-\phi),
 \end{equation}
 captures well the behaviour of the market price
 $I(t)$ prior to a
 crash or large correction at a time $\approx t_c$
 \cite{2,3}.
 
 There are many mechanisms known to generate
 log-periodicity
 \cite{1}. Let us stress that various dynamical
 mechanisms generate
 log-periodicity, without relying on a pre-existing
 discrete
 hierarchical structure. Thus, DSI may be produced
 dynamically (see
 in particular the recent nonlinear dynamical model
 introduced in
 \cite{4}) and does not need to be pre-determined
 by, e.g., a
 geometrical network. This is because there are many
 ways to break
 a symmetry, the subtlety here being to break it only
 partially.
 Thus, log-periodicity per se is not a signature of a
 critical
 point. Only within a well-defined model and
 mechanism can it be
 used as a potential signature. 
  
 Scale-invariance is a subgroup of a larger transformation group, 
 the conformal group. In any conformal field theory, the system shows
 invariance under
 translation, rotation, dilatation, and special
 conformal
 transformation \cite{5}. Conformal invariance is sufficient
 to determine
 the general form of simple correlation functions. It
 is known that
 generally the correlation functions in ordinary
 conformal field
 theories are in the form of scaling functions. In
 the logarithmic
 conformal field theories (LCFT), there exist at
 least one partner
 for a primary field. The general form of one-, two-,
 and
 three-point functions in LCFT's  have been obtained
 in [6-10].

 Recently, systems with discrete scale invariance
 have absorbed
 some interest. There does not exist, however, a
 complete study of
 general properties of such systems. In this paper we
 want to study
 the general form of correlation functions of systems
 with discrete
 scale invariance. In section 2, subgroups of
 the conformal transformations are investigated, which 
 contain discrete ({\em not full}) scale
 transformations. In section 3, the general form of
 the correlators of systems possessing discrete scale 
 invariance are obtained. In section 4 and 5, the same is 
 done for the systems possessing discrete scale invariance 
 plus translation- or special conformal 
 transformation-invariance, respectively.
 Both ordinary and logarithmic cases are studied.
 \section{Subgroups of the conformal group}
 Consider the transformations
 \begin{align}
 \label{s1}
 T_\alpha: z&\to z+\alpha,\qquad
 \hbox{translation}\nonumber\\
 S_\mu: z&\to e^\mu z,\qquad
 \hbox{scaling-rotation}\nonumber\\
 C_a: z&\to\frac{z}{1-az}\qquad \hbox{special
 conformal
 transformation}
 \end{align}
 in the complex plane. The group of transformations
 constructed by
 the above transformations, is that subgroup of the
 conformal
 group, the central extension of which is trivial. It
 is well-known
 that this group is isomorphic to SL$_2$($\mathbb
 C$). Let us call
 this group SCT. Throughout this section, we consider
 only those
 subgroups of SCT, which are \textit{complete}, by
 which it is
 meant that if a convergent sequence of
 group-elements of SCT are
 in the subgroup, the limit of that sequence is also
 in the
 subgroup.
 
 First consider subgroups of translations. Any
 subgroup of
 translations, consists of translations $T_\alpha$,
 with
 \begin{equation}\label{s2}
 \alpha=x\beta+y\beta',
 \end{equation}
 where $\beta$ and $\beta'$ are two fixed nonzero
 complex numbers,
 and $\beta/\beta'$ is not real. $x$ and $y$, each
 may take only
 real values, only integer values, or only zero.
 
 A similar argument holds for the subgroups of
 special conformal
 transformations, as these transformations are in
 fact translations
 of $1/z$.
 
 For the subgroups of scaling-rotations too, a
 similar argument
 holds. However, as a rotation by $2\pi$ ($S_\mu$
 with $\mu=2\pi
 i$) is equal to identity and hence should be in the
 subgroup, one
 concludes that any subgroup of scaling-rotations
 consists of the
 transformations $S_\mu$, with
 \begin{equation}\label{s3}
 \mu=x\nu+y\nu',
 \end{equation}
 where $\nu$ is pure imaginary, $\nu'$ is not pure
 imaginary, $\nu$
 and $\nu'$ are fixed. The values $x$ and $y$ can
 take, are as in
 the subgroups of translations, except that $x$ takes
 integer
 values only if $\nu$ is rational multiple of $2\pi
 i$. In this
 case, if $\nu=(2\pi i)n/m$, where $n$ and $m$ are
 integers prime
 with respect to each other and $m$ is positive, then
 those values
 of $x$ which result in distinct group-elements are
 integers from
 $0$ up to $m-1$. In this case, in fact one can use
 $\tilde\nu:=2\pi i/m$ instead of $\nu$. This means
 that subgroups
 of scaling-rotations always contain subgroups of
 pure rotations,
 which may be trivial, discrete, or the full rotation
 group.
 
 Next consider subgroups consisting of products of
 two of the above
 subgroups. These products are not direct products,
 since
 translations, scaling-rotations, and special
 conformal
 transformations do not commute with each other.
 Consider first a
 subgroup of SCT, consisting of transformations which
 are a
 scaling-rotation followed by a translation, where
 the translations
 are a subgroup of translations, and the
 scaling-rotations are a
 subgroup of scaling rotations:
 \begin{equation}\label{s4}
 z\to e^\mu z+\alpha,
 \end{equation}
 where $\alpha$ takes value on a lattice, (\ref{s2}),
 and $\mu$
 takes value on another lattice, (\ref{s3}). (Each of
 these
 lattices may be degenerate, that is, a collection of
 lines, the
 whole plane, just one point, etc.) This set of
 transformations is
 a subgroup, if and only if the lattice of the
 possible values of
 $\alpha$ is invariant under the action of
 scaling-rotations which
 are in the set. The only possible ways (apart the
 trivial cases
 that the scaling-rotations or translations consist
 of only the
 identity element) are the following.
 \begin{itemize}
 \item[i)] a rectangular lattice of translations,
 with
 scaling-rotations only rotations by integer
 multiples of $\pi/2$.
 \item[ii)] a triangular lattice of translations,
 with
 scaling-rotations only rotations by integer
 multiples of $\pi/3$
 or $2\pi/3$.
 \item[iii)] Any lattice of translations, with
 scaling-rotations
 only rotations by integer multiples of $\pi$.
 \item[iv)] a degenerate lattice of translations,
 consisting of
 one continuous line, with scaling-rotations only a
 subgroup (any
 subgroup) of scalings, or the product of a subgroup
 of scalings
 and rotations by integers multiples of $\pi$.
 \item[v)] the full translations with any subgroup of
 scaling-rotations.
 \end{itemize}
 Again, a similar argument holds for subgroups of
 SCT, consisting
 of transformations which are a scaling-rotation
 followed by a
 special conformal transformation, where the special
 conformal
 translations are a subgroup of special conformal
 transformations,
 and the scaling-rotations are a subgroup of scaling
 rotations:
 \begin{equation}\label{s5}
 z^{-1}\to  e^{-\mu}z^{-1}-a.
 \end{equation}
 
 A set of transformations consisting of a translation
 followed by a
 special conformal transformation, where the special
 conformal
 translations are a subgroup of special conformal
 transformations,
 and the translations are a subgroup of translations,
 cannot be a
 subgroup, unless the only translation is identity,
 or the only
 special conformal transformation is the identity. To
 show this,
 one notices that the product $T_\alpha S_a$, can be
 equal to
 $C_{a'}T_{\alpha'}$ for some $\alpha'$ and $a'$,
 only if
 $a\alpha=0$ or $a\alpha=2$. But this criterion is
 necessary for
 the above-mentioned set to be a subgroup (the
 multiplication be
 closed). If $a\alpha=0$ for all translations and
 special conformal
 translations in the set, then at least on of these
 subgroups must
 contain no element apart from the identity. If
 $a\alpha=2$ for
 some $a$ and $\alpha$, one notes that if the
 above-mentioned set
 contains $T_\alpha$, it should also contain
 $T_{2\alpha}$, and
 obviously $2a\alpha\ne 2$.
 
 Finally, let us consider a subgroup G of SCT,
 containing
 nonidentity translations, scaling-rotations, and
 special conformal
 transformation. First, assume also that $G$ contains
 a
 scaling-rotation which is not a pure-rotation, that
 is, some
 $S_\mu$, with $\mu$ not pure imaginary. The above
 arguments show
 that G must contain a subgroup of continuous
 translations at least
 in one direction, and a subgroup of continuous
 special conformal
 translations at least in one dimension. The
 generators of these
 two one-parameter transformations are
 \begin{equation}\label{s6}
 g_1=^{i\theta}L_{-1}+e^{-i\theta}\bar L_{-1},
 \end{equation}
 and
 \begin{equation}\label{s7}
 g_2=e^{i\zeta}L_1+e^{-i\zeta}\bar L_1,
 \end{equation}
 for some $\theta$ and $\zeta$. Using the commutation
 relation of
 the generators, it is seen that G contains another
 one-parameter
 continuous subgroup, with the generator
 \begin{equation}\label{s8}
 g_3=e^{i(\theta+\zeta)}L_0+e^{-i(\theta+\zeta)}\bar
 L_0.
 \end{equation}
 So G contains a continuous subgroup of
 scaling-rotations.
 Combining this with the above arguments, it is seen
 that either
 \begin{itemize}
 \item[i)] G consists of If transformations generated
 by $g_1$,
 $g_2$, and
 $L_0+\bar L_0$, where $\alpha+\zeta$ is an integer
 multiple of
 $\pi$, or,
 \item[ii)] G is equal to SCT.
 \end{itemize}
 Second, assume that the only scaling-rotations
 contained in G are
 pure rotations. It can be shown that G cannot
 contain both
 nontrivial translations and nontrivial special
 conformal
 transformations. The argument is similar to that
 used to prove
 that a set containing translations followed by
 special conformal
 transformations cannot be a subgroup of SCT. The
 difference is
 that now, For any $T_\alpha$ and $S_a$ in G, there
 should be some
 $T_{\alpha'}$ and $S_{a'}$ in G, so that
 $T_{-\alpha'}S_{-a'}T_\alpha S_a$ be a pure
 rotation.
 
 The above arguments can be stated for the conformal
 group in one
 dimension as well. In one dimension, SCT is still
 constructed by
 the transformations $T_\alpha$, $S_\mu$, and $C_a$.
 But now the
 parameters $\alpha$, $\mu$, and $a$ are real, the
 scaling-rotations are in fact only scalings, and SCT
 is isomorphic
 to SL$_2$($\mathbb R$). Arguments similar to the
 above, then show
 that there are subgroups of SCT consisting of
 \begin{itemize}
 \item all translations, or discrete translations
 \item all special conformal transformations, or
 discrete special
 conformal transformations
 \item all scalings, or discrete scalings
 \item all translations, and all scalings or discrete
 scalings
 \item all special conformal translations, and all
 scalings or
 discrete scalings
 \end{itemize}
 \section{Systems with discrete scale invariance}
 Consider a one-dimensional system with discrete
 scale invariance.
 A quasi-primary field transforms under the scaling
 as:
 \begin{equation}\label{1}
 \phi(x)\to e^{\Delta\mu_1}\phi(e^{\mu_1} x).
 \end{equation}
 Invariance under discrete scaling gives
 \begin{equation}\label{2}
 \langle\phi(x)\rangle=e^{\Delta\mu_1}
 \langle\phi(e^{\mu_1}x)\rangle.
 \end{equation}
 Defining
 \begin{equation}\label{2p}
 g(x):=x^\Delta\langle\phi(x)\rangle,
 \end{equation}
 one arrives at
 \begin{equation}\label{3}
 g(e^{\mu_1}x)=g(x).
 \end{equation}
 This shows that $g$ is periodic with respect to
 $\log x$, with the
 period $\mu_1$. Hence it has a Fourier series:
 \begin{equation}\label{5}
 g(x)=\sum_{n=-\infty}^{\infty} C_n\exp\left({{2\pi n
 i\ln x}
 \over{\mu_1}}\right),
 \end{equation}
 from which one can obtain the one-point function
 $\langle\phi(x)\rangle$. The expectation value of
 the field $\phi$
 is real, iff $C_n=\bar C_{-n}$. In this case, one
 arrives at
 \begin{equation}\label{6}
 \langle\phi(x)\rangle=x^{-\Delta
 }\sum_{n=0}^{\infty}
 r_n\cos\left({{2\pi n\ln
 x}\over{\mu_1}}+\theta_n\right),
 \end{equation}
 where $r_n$ and $\theta_n$ are defined through
 $C_n=:r_n\exp(i\theta_n)$.
 
 The two-point function of two quasi-primary fields
 can be obtained
 similarly. Defining
 \begin{equation}\label{7}
 g(x_1,x_2):=x_1^{\Delta_1}x_2^{\Delta_2}
 \langle\phi_1(x_1)\phi_2(x_2)\rangle,
 \end{equation}
 and exploiting the discrete scale invariance, it is
 seen that
 \begin{equation}\label{8}
 g(e^{\mu_1}x_1,e^{\mu_1}x_2)=g(x_1,x_2),
 \end{equation}
 from which one arrives at
 \begin{equation}\label{9}
g(x_1,x_2)=h\left({{x_2}\over{x_1}}\right)P\left({{\ln
 x_1}\over{\mu_1}}\right),
 \end{equation}
 where $h$ is an arbitrary function, and $P$ is a
 periodic function
 with period one. One can then expand this function
 as a Fourier
 series, and obtain the two-point function as
 \begin{equation}\label{10}
 \langle\phi_1(x_1)\phi_2(x_2)\rangle=x_1^{-\Delta_1}
 x_2^{-\Delta_2}h\left({{x_2}\over{x_1}}\right)
 \sum_{n=-\infty}^\infty C_n\exp\left({{2\pi n i\ln
 x_1}
 \over{\mu_1}}\right),
 \end{equation}
 where the constants $C_n$'s and the function $h$ are
 arbitrary.
 
 This argument can be extended to more-point
 functions of
 quasi-primary fields, and one arrives at
 \begin{equation}\label{11}
\langle\phi_1(x_1)\cdots\phi_k(x_k)\rangle=x_1^{-\Delta_1}\cdots
x_k^{-\Delta_k}h\left({{x_2}\over{x_1}},\dots,{{x_k}\over{x_{k-1}}}
 \right)\sum_{n=-\infty}^\infty C_n\exp\left({{2\pi n
 i\ln x_1}
 \over{\mu_1}}\right).
 \end{equation}
 
 Now let us assume a logarithmic partner $\psi$ for
 the
 quasi-primary field $\phi$. This means that under
 discrete scale
 transformation, $\psi$ transforms like
 \begin{equation}\label{12}
 \psi(x)\to e^{\Delta\mu_1}[\psi(e^{\mu_1} x)+\mu_1
 \phi(e^{\mu_1}
 x)].
 \end{equation}
 This is nothing by the formal derivative of \Ref{1}
 with respect
 to $\Delta$, if one considers $\psi$ as the formal
 derivative of
 $\phi$ with respect to $\Delta$. So, in light of
 [8],
 it is not surprising that the expectation of $\psi$
 is just the
 formal derivative of that of $\phi$:
 \begin{equation}\label{13}
\langle\psi(x)\rangle=x^{-\Delta}\sum_{n=-\infty}^{\infty}
 (C'_n-C_n\ln x)\exp\left({{2\pi n i\ln x}
 \over{\mu_1}}\right),
 \end{equation}
 where $C'_n$ is the formal derivative of $C_n$ with
 respect to
 $\Delta$. This can be readily extended to more-point
 functions:
 Changing any quasi-primary field in the left-hand
 side of
 \Ref{11}, into its logarithmic partner, one has to
 differentiate
 formally the right-hand side with respect to the
 weight of that
 field, treating arbitrary constants and functions as
 entities
 depending on that weight.
 
 Finally, the above arguments for obtaining the
 correlators of a
 one-dimensional system with discrete scale
 invariance, can be
 easily extended to two-dimensional system with
 rotational
 invariance and discrete scale invariance. Here, a
 quasi-primary
 field $\phi$ transforms as
 \begin{align}\label{1p}
 \phi(z)&\to
 e^{i(\Delta-\bar\Delta)\theta}\phi(e^{i\theta}
 z),\qquad z\to e^{i\theta}z,\nonumber\\
 \phi(z)&\to
 e^{(\Delta+\bar\Delta)\mu_1}\phi(e^{\mu_1} z),\qquad
 z\to e^{\mu_1}z.
 \end{align}
 (It is not assumed that $\phi$ is holomorphic. Its
 dependence on
 $z$ is a short-hand notation for its dependence on
 the real-part
 and the imaginary-part of $z$, or its dependence on
 $z$ and $\bar
 z$.) Now, let us find the expectation of $\phi$.
 Defining
 \begin{equation}\label{15}
 g(z):=z^\Delta\bar
 z^{\bar\Delta}\langle\phi(z)\rangle,
 \end{equation}
 it is seen that $g$ depends on only $|z|$, and that
 $g$ is
 periodic with respect to $\ln|z|$, with the period
 $\mu_1$. So,
 \begin{equation}\label{16}
 \langle\phi(z)\rangle=z^{-\Delta}\bar
 z^{-\bar\Delta}
 \sum_{n=-\infty}^{\infty} C_n \exp\left({{2\pi n
 i\ln|z|}\over{\mu_1}}\right).
 \end{equation}
 A similar reasoning leads to
 \begin{align}\label{17}
\langle\phi_1(z_1)\cdots\phi_k(z_k)\rangle=&z_1^{-\Delta_1}
 \cdots
 z_k^{-\Delta_k}\bar z_1^{-\bar\Delta_1}\cdots \bar
 z_k^{-\bar\Delta_k}
h\left({{z_2}\over{z_1}},\dots,{{z_k}\over{z_{k-1}}}\right)
 \nonumber\\ &\times\sum_{n=-\infty}^\infty
 C_n\exp\left({{2\pi n
 i\ln|z_1|} \over{\mu_1}}\right),
 \end{align}
 for the $k$-point function of $k$ quasi-primary
 fields.
 Differentiating this formally, with respect to
 appropriate
 weights, one arrives at the correlators containing
 logarithmic
 partners as well.
 \section{Discrete scale invariance plus translation
 invariance}
 First consider a one-dimensional system with
 discrete scale
 invariance as well as translation invariance. A
 quasi-primary
 field $\phi$, is transformed under discrete scale
 transformation
 like \Ref{1}, and is transformed under translation
 $x\to x+\alpha$
 as
 \begin{equation}\label{18}
 \phi(x)\to\phi(x+\alpha).
 \end{equation}
 Translation invariance results in
 \begin{equation}\label{19}
 \langle\phi(x+a)\rangle=\langle\phi(x)\rangle,
 \end{equation}
 or
 \begin{equation}\label{20}
 \langle\phi(x)\rangle=C,
 \end{equation}
 where $C$ is a constant. Discrete scale invariance,
 then gives
 \begin{equation}\label{21}
 (e^{\Delta\mu_1} -1)C=0.
 \end{equation}
 This means that the one-point function is nonzero,
 only if for
 fields with the weights
 \begin{equation}\label{22}
 \Delta={{2\pi k i}\over{\mu_1}},
 \end{equation}
 where $k$ is an integer. If there exists a
 logarithmic partner for
 $\phi$, then translational invariance gives
 \begin{align}\label{23}
 \langle\phi(x)\rangle&=C_1,\nonumber\\
 \langle\psi(x)\rangle&=C_2.
 \end{align}
 Here $\psi$ is the logarithmic partner of $\psi$,
 being
 transformed under discrete scale invariance like
 \Ref{12}, and
 behaving under translation like \Ref{18}, with
 $\phi$ substituted
 with $\psi$. Imposing discrete scale invariance, it
 is seen that
 the one-point functions are zero, unless
 $e^{\Delta\mu_1}=1$, and
 in this case $C_1$ is zero. The constraint on the
 weight is the
 same as \Ref{22}. So,
 \begin{align}\label{24}
 \langle\phi(x)\rangle&=0,\nonumber\\
 \langle\psi(x)\rangle&=C_2,\nonumber\\
 e^{\Delta\mu_1}&=1.
 \end{align}
 
 The more-point functions of quasi-primary fields,
 can be obtained
 similarly. Translation invariance makes the
 $k$-point function a
 function of $(k-1)$ independent differences of the
 $k$
 coordinates. Defining
 \begin{align}\label{25}
 g(x_2-x_1,\dots,x_k-x_{k-1}):=&(x_1-x_k)^{\Delta_1}
(x_2-x_1)^{\Delta_2}\cdots(x_k-x_{k-1})^{\Delta_k}\nonumber\\
 &\times\langle\phi(x_1)\cdots\phi_k(x_k)\rangle,
 \end{align}
 it is seen that
 \begin{equation}\label{26}
g(x_2-x_1,\dots,x_k-x_{k-1})=h\left({{x_3-x_2}\over{x_2-x_1}},
 \dots,{{x_k-x_{k_1}}\over{x_{k-1}-x_{k-2}}}\right)
 P\left[{{\ln(x_2-x_1)}\over{\mu_1}}\right],
 \end{equation}
 where $h$ is an arbitrary function and $P$ is
 periodic with the
 period one. From this, the $k$-point function is
 obtained as
 \begin{align}\label{27}
\langle\phi(x_1)\cdots\phi_k(x_k)\rangle=&(x_1-x_k)^{-\Delta_1}
(x_2-x_1)^{-\Delta_2}\cdots(x_k-x_{k-1})^{-\Delta_k}\nonumber\\
 &\times h\left({{x_3-x_2}\over{x_2-x_1}},\dots,
{{x_k-x_{k_1}}\over{x_{k-1}-x_{k-2}}}\right)\nonumber\\&\times
 \sum_{n=-\infty}^\infty C_n\exp\left[{{2\pi n
 i\ln(x_2-x_1)}
 \over{\mu_1}}\right].
 \end{align}
 Correlators containing logarithmic parts, can be
 obtained by
 formal differentiation with respect to appropriate
 weights. The
 one-point function is exceptional, as it is zero
 unless the weight
 is one of the members of a discrete set, eq.
 \Ref{22}. This case
 was discussed previously.
 
 Finally, for a two-dimensional system, a
 quasi-primary field
 $\phi$ is transformed under rotation and discrete
 scaling like
 \Ref{1p}, and under translation like the obvious
 generalization of
 \Ref{18}. Again, translation invariance makes the
 one-point
 function of the quasi-primary field $\phi$
 independent of $z$:
 \begin{equation}\label{28}
 \langle\phi(z)\rangle=C.
 \end{equation}
 Rotation invariance and discrete scale invarinace,
 then result in
 \begin{align}\label{29}
 e^{i(\Delta-\bar\Delta)\theta}C&=C,\nonumber\\
 e^{(\Delta+\bar\Delta)\mu_1}C&=C,
 \end{align}
 respectively. From the first equation, $C$ may be
 nonzero only if
 $\Delta=\bar\Delta$. Then the second equation shows
 that $C$ may
 be nonzero only if
 \begin{equation}\label{30}
 \Delta=\bar\Delta={{\pi k i}\over{\mu_1}},
 \end{equation}
 where $k$ is an integer. If $\phi$ has a logarithmic
 partner
 $\psi$, then there may be nonzero one-point
 functions only as
 \begin{align}\label{31}
 \langle\phi(z)\rangle&=0,\nonumber\\
 \langle\psi(z)\rangle&=C_2,\nonumber\\
 e^{(\Delta+\bar\Delta)\mu_1}&=1.
 \end{align}
 
 $k$-point functions of $k$ quasi-primary fields are
 also similarly
 obtained:
 \begin{align}\label{32}
\langle\phi(z_1)\cdots\phi_k(z_k)\rangle=&(z_1-z_k)^{-\Delta_1}
(z_2-z_1)^{-\Delta_2}\cdots(z_k-z_{k-1})^{-\Delta_k}\nonumber\\
 &\times(\bar z_1-\bar z_k)^{-\bar\Delta_1} (\bar
 z_2-\bar
 z_1)^{-\bar\Delta_2} \cdots(\bar z_k-\bar
 z_{k-1})^{-\bar\Delta_k}
 \nonumber\\&\times
 h\left({{z_3-z_2}\over{z_2-z_1}},\dots,
{{z_k-z_{k_1}}\over{z_{k-1}-z_{k-2}}}\right)\nonumber\\&\times
 \sum_{n=-\infty}^\infty C_n\exp\left[{{2\pi n
 i\ln|z_2-z_1|}
 \over{\mu_1}}\right].
 \end{align}
 More-than-one-point functions containing logarithmic
 partners, are
 also obtained by formal differentiation of the
 above, with respect
 to appropriate weights.
 \section{Discrete scale invariance plus special
 conformal
 transformation invariance} Now, let's consider the
 systems
 invariant under discrete scale transformation
 together with
 continous special conformal transformation. First
 consider
 one-dimensional systems. A quasi-primary field
 $\phi$, is
 transformed under special conformal transformation
 like
 \begin{equation}\label{33}
 \phi(x)\to{1\over{(1-ax)^{2\Delta}}}
 \phi\left({x\over{1-ax}}\right).
 \end{equation}
 Defining a field $\tilde\phi$ through
 \begin{equation}\label{34}
 \tilde\phi(x):=x^{-2\Delta}\phi\left({1\over
 x}\right),
 \end{equation}
 it is seen that under $C_a$, this field is
 transformed like
 \begin{equation}\label{35}
 \tilde\phi(x)\to\tilde\phi(x-a).
 \end{equation}
 It is also seen that under discrete scaling,
 $\tilde\phi$ is
 transformed as
 \begin{equation}\label{36}
 \tilde\phi(x)\to
 e^{-\Delta\mu_1}\tilde\phi(e^{-\mu_1}x).
 \end{equation}
 These show that corresponding to the quasi-primary
 field $\phi$,
 there exists another field $\tilde\phi$ such that
 the actions of
 special conformal transformation and discrete
 scaling on which are
 like translation and discrete scaling, respectively.
 So
 correlators containing these new fields are like
 those obtained in
 the previous section. From these, it is easy to
 obtain correlators
 containing the quasi-primary fields. If $\phi$ has a
 logarithmic
 partner $\psi$, then define the field $\tilde\psi$
 through
 \begin{equation}\label{37}
 \tilde\psi(x):=x^{-2\Delta}\left[\psi\left({1\over
 x}\right)
 -2\Delta\ln x\,\phi\left({1\over x}\right)\right].
 \end{equation}
 This is in fact the formal derivative of
 $\tilde\phi$ with respect
 to $\Delta$. Correlators containing $\tilde\phi$ and
 $\tilde\psi$
 are of the general forms obtained in the previous
 section. From
 these, it is easy to obtain correlators containing
 $\phi$ and
 $\psi$, or several fields like them.
 
 For two-dimensional systems, one defines
 $\tilde\phi$
 corresponding to a quasi-primary field $\phi$ like
 \begin{equation}\label{38}
 \tilde\phi(z):=z^{-2\Delta}\bar
 z^{-2\bar\Delta}\phi\left({1\over
 z}\right).
 \end{equation}
 So correlators containing $\tilde\phi$'s are of the
 general form
 obtained in the previous section. Correlators
 containing
 logarithmic partners can be obtained through
 differentiation with
 respect to appropriate weights.

 \end{document}